# Observation of bosonic condensation in a hybrid monolayer MoSe$_2$-GaAs microcavity

Max Waldherr[1]†, Nils Lundt[1]†, Martin Klaas[1]†, Simon Betzold[1], Matthias Wurdack[1], Vasilij Baumann[1], Eliezer Estrecho[2], Anton Nalitov[3,4,5], Evgenia Cherotchenko[4,5], Hui Cai[6], Elena A. Ostrovskaya[2,7], Alexey V. Kavokin[5,8,9], Sefaattin Tongay[6], Sebastian Klembt[1], Sven Höfling[1,10] and Christian Schneider[1]*

[1]*Technische Physik and Wilhelm-Conrad-Röntgen-Research Center for Complex Material Systems, Universität Würzburg, D-97074 Würzburg, Am Hubland, Germany.*

[2]*Nonlinear Physics Centre, Research School of Physics and Engineering, Australian National University, Canberra ACT 2601, Australia*

[3]*Science Institute, University of Iceland, Dunhagi 3, IS-107, Reykjavik, Iceland*

[4]*ITMO University, St. Petersburg 197101, Russia*

[5]*Physics and Astronomy School, University of Southampton, Highfield, Southampton, SO171BJ, UK*

[6]*School for Engineering of Matter, Transport, and Energy, Arizona State University, Tempe, Arizona 85287, USA*

[7]*ARC Centre of Excellence in Future Low-Energy Electronics Technologies*

[8]*SPIN-CNR, Viale del Politecnico 1, I-00133 Rome, Italy*

[9]*Spin Optics Laboratory, St-Petersburg State University, 1, Ulianovskaya, 194021, Russia*

[10]*SUPA, School of Physics and Astronomy, University of St. Andrews, St. Andrews KY 16 9SS, United Kingdom*

†These authors contributed equally to this work
*Corresponding author. Email: Christian.Schneider@physik.uni-wuerzburg.de

**Condensation of bosons into a macroscopic quantum state belongs to the most intriguing phenomena in nature. It was first realized in quantum gases of ultra-cold atoms, but more recently became accessible in open-dissipative, exciton-based solid-state systems at elevated temperatures. Semiconducting monolayer crystals have emerged as a new platform for studies of strongly bound excitons in ultimately thin materials. Here, we demonstrate the formation of a bosonic condensate driven by excitons hosted in an atomically thin layer of MoSe$_2$, strongly coupled to light in a solid-state resonator. The structure is operated in the regime of collective strong coupling, giving rise to hybrid exciton-polariton modes composed of a Tamm-plasmon resonance, GaAs quantum well excitons and two-dimensional excitons confined in a monolayer of MoSe$_2$. Polariton condensation in a monolayer crystal manifests by a superlinear increase of emission intensity from the hybrid polariton mode at injection powers as low as 4.8 pJ/pulse, as well as its density-dependent blueshift and a dramatic collapse of the emission linewidth as a hallmark of temporal coherence. Importantly, we observe a significant spin-polarization in the injected polariton condensate, a fingerprint of the core property of monolayer excitons subject to spin-valley locking. The observed effects clearly underpin the perspective of building novel highly non-linear valleytronic devices based on light-matter fluids, coherent bosonic light sources based on atomically thin materials, and paves the way towards studying materials with unconventional topological properties in the framework of bosonic condensation.**

Bosonic condensation is an intriguing phenomenon, which describes the collective collapse of quantum particles into a single macroscopic and coherent quantum state. For a long time, experiments devoted to bosonic condensation have been reserved for ultra-cold atoms[1,2], but more recently became accessible in open-dissipative solid state systems[3] at elevated temperatures. A prime candidate to observe bosonic condensation in solids are exciton-polaritons, which are bosonic quasi-particles resulting from strong light-matter coupling in microcavities with embedded materials that are characterized by a large dipole oscillator strength[4,5]. These composite particles possess a variety of very appealing physical properties, particularly prominent at larger densities. They are bosons with a very low and tunable effective mass and are therefore almost ideally suited for studies of Bose-Einstein condensation phenomena[3,6] at elevated temperatures[7,8].

The intrinsic properties of exciton-polaritons critically dependent on properties of the matter excitations. Atomically thin monolayers of transition metal dichalcogenides (TMDCs) have emerged as a new material platform with highly interesting excitonic properties: The materials are mono-atomically thin, thus composing the ultimate physical limit for a system to host collective electronic excitations. They are highly non-linear, and the chiral exciton properties are uniquely linked to the valley degree of freedom, i.e. excitons emerging from the direct band transition at the K-point (K'-point) possess a pseudospin projection to the structure axis of +1 (-1). Locking of spin- and valley index directly protects valley excitons from fast spin-relaxation for small exciton momenta[9–11], and more importantly, it allows to interlink excitonic propagation with its chirality, driven by the Berry curvature of the valleys[12]. This new effect paved the way to the new research area of valleytronics[12–14].

Valleytronics with excitons intrinsically suffers from the very short diffusion length of the optically addressable quasi-particles, which intrinsically limits the zoo of observable phenomena, or makes them at least very hard to detect with conventional experimental methods. This includes the possible manifestation of chiral excitonic currents on the edge of carefully prepared twisted bilayers[15], the interplay between long-range order[3] and valley physics[12] and the interplay between superconductivity and superfluidity[16,17] which has been predicted in such systems. Cavity exciton-polaritons can, in principle, provide a feasible solution to this roadblock, as the expansion of a polariton cloud in a microcavity and the built-up of its coherence is known to be very fast. In this spirit, it has been shown that the valley selective strong coupling regime with excitons in atomic monolayers of $MoS_2$[18], $MoSe_2$[19,20], $WS_2$[21] and a microcavity resonance in the linear (low density) regime is accessible.

However, pronounced polariton expansion over macroscopic distances[22], the formation of robust spin patterns[23], and the emergence of topological excitations[24] are expected in the non-linear (high density) regime of bosonic condensation. This regime is anticipated in microcavities with embedded TMDC monolayers, but has not been accessible so far due to experimental limitations.

Here, we utilize the collective coupling[25] of excitons in GaAs quantum wells and $MoSe_2$ to a joint photonic resonance, combining efficient polariton energy relaxation with robust preservation of an optically induced spin-polarization. The hybridization leads to a longer radiative lifetime of exciton-polaritons and their reservoirs[26], and particularly, it facilitates the build-up of the critical population for bosonic stimulation in the ground state. We detect the condensation of exciton-polaritons into the lowest, quantized energy state of the hybrid mode by measuring its population and infer the coherence by tracing the spectral width of the emitted light. In the regime of hybrid polariton condensation, we observe clear indications of interactions between polaritons and excitons in the optically injected reservoir by studying the power-dependent energy shift of the mode and confirm that the valley index of the monolayer remains addressable.

A sketch of the studied photonic microstructure, giving rise to so-called Tamm-plasmon polariton resonances[27,28], is depicted in Fig. 1a. The structure is similar to the one described in Ref. [29]: It consists of an AlAs/AlGaAs distributed Bragg reflector (DBR) (30 pairs), which is characterized spectrally by its stopband ranging from 710 nm to 790 nm (see Ref. [29]), with reflectivity up to 99.9 % between 740 nm (1.675 eV) and 765 nm (1.621 eV) at 10 K. The AlAs/AlGaAs Bragg stack, which has been grown by gas source molecular beam epitaxy, is topped with a 112 nm thick AlAs layer with four embedded GaAs quantum wells (QWs). A layer of GaInP caps the AlAs layer. A single monolayer of $MoSe_2$, mechanically exfoliated via commercial adhesive tape from a bulk crystal was transferred onto the top GaInP layer with a polymer stamp[30].

The full cavity device is completed by capping the monolayer with an 80 nm thick layer of polymethyl methacrylate (PMMA) and a 60 nm thick gold layer, such that it promotes an optical resonance with a field antinode both at the position of the monolayer and at the stack of GaAs QWs (Fig. 1b), and yields an estimated Q-factor of 650. We characterized the absorption of our embedded material in a high resolution measurement prior to completion of our microcavity at a sample temperature of 10 K (Fig. 1c). There, we traced out two absorption resonances, which we attribute to the free exciton in the $MoSe_2$ monolayer, as well as the GaAs QWs.

## Results

**Hybrid $MoSe_2$-GaAs exciton polariton mode**

In order to confirm the emergence of a hybrid polariton mode in our device, we first study the dispersion relation of the bare GaAs QW exciton-polaritons by recording the emission at a position near the monolayer, which is depicted in Fig. 2a. The spectra in the waterfall representation in Fig. 2a were extracted from angle resolved emission spectra, which we recorded in a modified micro-photoluminescence setup in the far field imaging configuration (see Materials and Methods). We can clearly construct the dispersion relation of the QW exciton-polaritons with a mass of $4\times10^{-5}$ times the free electron mass, a Rabi splitting of 9.2 meV and a positive exciton-photon detuning of 5.9 meV. These observations are fully consistent with previous results reported in a similar device[29].

Next, we record the spectral emission signatures of our sample at the position of the monolayer. Due to the small size of the flake (approx. 1.7 µm in diameter, see below), along with the emission from the monolayer, we also collect signal from its periphery due to the lateral diffusion of GaAs quantum well excitons. The corresponding spectra shown in Fig. 2c are composed of a high-energy dispersive signal, which follows precisely the dispersion relation of the bare III-V GaAs exciton polariton in Fig. 2a and a second, red-shifted signal of a dispersion-less resonance spread by 2.5 µm$^{-1}$ in the momentum space. That signal is consistent with the formation of a spatially confined hybrid exciton-polariton mode in the collective strong coupling regime between excitons in the $MoSe_2$ monolayer and GaAs quantum well, respectively, and the cavity mode. The finite size of the monolayer induces a strong quantization of the hybrid mode yielding a confinement-induced blueshift of 1.21 meV with respect to the energy minimum of the hybrid mode at the lowest excitation power. This dispersion relation was calculated assuming a coupling strength of 20 meV between the monolayer exciton and the Tamm-resonance, that is based on our previous findings[29]. By treating the monolayer as a finite potential well with the confinement depth of 2.38 meV, which is given by the difference between the ground state of the GaAs polariton dispersion outside the monolayer and the minimum of the hybrid polariton dispersion, we deduce the lateral extent of the monolayer of approximately 1.7 µm.

**Condensation of Polaritons**

The formation of a condensate of exciton-polaritons emerging in our hybrid mode can be visualized in power dependent experiments, which are carried out at a sample temperature of 4.2 K. In Fig. 3a-d, we plot far-field spectra recorded from the position of the monolayer at various pump powers. Exciton-polaritons are injected in the system by 2 ps long laser pulses (82 MHz repetition rate). The wavelength is tuned approximately to the energy of the upper hybrid polariton branch at 741 nm. Here, we observe that with growing pump powers the spectrum is progressively dominated by the identified hybrid mode. At the same time, the hybrid mode undergoes a distinct energy blueshift and its spectral width narrows. In Fig. 3e, we provide a more detailed analysis of this primary emission feature. From the plotted input-output curve, we can deduce a clear threshold behavior at pump energies as low as 4.8 pJ/pulse (grey shaded area), which is accompanied by a rapid drop in the polariton linewidth from 2.1 eV down to 0.7 meV. This drop in the linewidth is a strong evidence of the onset of temporal coherence in systems emerging at the transition from a thermal to a coherent state[31]. One crucial difference between polariton condensates and classical laser modes is manifested by the excitonic component of cavity polaritons, which governs the polariton interactions with uncoupled excitons and other polaritons. These interactions induce the blueshift of the polariton mode, which is plotted in Fig. 3f as a function of the energy per injection pulse. The inset depicts the energy shift as a function of the polariton occupancy in the hybrid mode, which is calculated by normalizing the emission intensity by the intensity at the threshold. Below the condensation threshold, the hybrid mode experiences an approximately linear blueshift with increasing pump power on the order of 3 meV, which we attribute primarily to the interaction between the hybrid polaritons and excitons in the non-resonantly driven reservoirs which are continuously built up. As we cross the threshold, the mode continues to blueshift by approximately 300 µeV (also see inset), yet the slope of the curve changes which reflects the modified growth rates of the two reservoirs due to formation of the polariton condensate (see apendix). This effect can be accounted for by using an analytical model which constitutes a set of semi-classical Boltzmann rate equations accounting for the two excitonic reservoirs (see appendix) and produces a good fit to our data as shown in Fig. 3f.

We note that, at the largest pump powers, we observe a second deviation from the linear increase of intensity above the condensation threshold (accompanied by a drop in the emission linewidth). This deviation is in general agreement with the strong to weak coupling laser transition for the GaAs-polaritons, which was observed in a comparative measurement in the vicinity of the MoSe$_2$ monolayer (see appendix).

**Valley polarization**

Finally, we address the question of whether the valley index of monolayer excitons can be controlled and preserved in a condensate of hybrid exciton polaritons. Therefore, we drive the system by a circularly polarized injection laser, to inject excitons predominantly in one valley of the embedded monolayer. As our laser is injecting quasi-particles into the upper hybrid mode, we expect that we predominantly create polaritons tagged by one valley index. This is reflected by the circular polarization of the condensate shown in Fig. 4a and 4b which clearly retains the polarization of the pump laser to a significant degree. The degree of circular polarization (DOCP) is calculated via $(N^{\pm} - N^{\mp})/(N^{\pm} + N^{\mp})$ and yields a value as high as 17.9 % for σ$^+$-pumping (16.4 % for σ$^-$-pumping) for an average polariton population at the pumping power of 10 $P_{th}$. The degree of polarization emitted from the hybrid mode is strongly reduced to approximately 9% in the linear regime (Fig. 4a and 4b). As a reference, the bare GaAs exciton polaritons show a reduced circular polarization in the linear (low-density) regime (recorded at the comparable laser-exciton detuning of 15 meV) of approx. 7 %, which does not substantially change with the pump power (appendix).

In order to provide a model to support our findings, we consider two spin-polarized excitonic reservoirs (in the GaAs QW and the monolayer), which are in turn generated by a circularly polarized optical excitation. This system can be described with a set of semi-classical Boltzmann equations, similar to those introduced in the appendix, but now also accounting for the spin-polarization of the excitonic reservoirs and condensed polaritons:

$$\frac{dN^\pm}{dt} = \left(W_1 n_1^\pm + W_2 n_2^\pm\right)(N^\pm + 1) - \frac{N^\pm}{\tau} - \frac{N^\pm - N^\mp}{\tau_s} \quad (1)$$

$$\frac{dn_i^\pm}{dt} = P_i^\pm - W_{1(2)}^\pm n_i^\pm (N^\pm + 1) - \frac{n_i^\pm}{\tau_i} - \frac{n_i^\pm - n_i^\mp}{\tau_{s,i}} \quad (2)$$

Here $N^\pm$ and $n_i^\pm$ are the condensate and the reservoir densities in the two spin components, $W_i$ are the spontaneous scattering rates from the two reservoirs, $\tau$, $\tau_i$, $\tau_s$ and $\tau_{s,i}$ are the lifetimes and spin relaxation times of the condensate and the reservoirs, while $P_i$ are the pumping rates of the reservoirs, where $i = 1$ corresponds to the QW and $i = 2$ to the monolayer. We further assume that both reservoirs are created with a circularly polarized pumping:

$$P_i^- = 0; \qquad P_i^+ = g_i P, \quad (3)$$

where $g_1 = 1$ and $g_2 = 1(0)$ in the hybrid(QW) cavity case. We assume that the pumping is equally efficient in the monolayer and GaAs QW.

In Fig. 4c we plot the power evolution of the degree of circular polarization in the hybrid polariton system (red), in comparison with the pure GaAs system (blue). We note that the degree of circular polarization arises from a subtle interplay between spin-valley relaxation and bosonic condensation in the system. While, in the case of hybrid polaritons, the polariton pseudospin is substantially better protected from depolarization by the effects of spin-valley locking, this effect is strongly enhanced by the bosonic amplification, which speeds up the relaxation dynamics from the reservoir, as reflected in our experiment.

## Conclusion

In conclusion, we studied the density dependence of hybrid exciton-polaritons arising in a microcavity with four embedded GaAs QWs and a single monolayer of MoSe$_2$. The formation of a condensate of exciton polaritons is clearly manifested by the strong non-linearity in the input-output characteristics occurring at pump powers as low as 4.8 pJ/pulse, the collapse of the emission linewidth of the hybrid mode and characteristic change of the blueshift above threshold signifying macroscopic occupation of the polariton mode. We demonstrate the effect of spin-valley locking in our condensate, a substantial feature inherent from the atomic monolayer, which paves the way to uniquely study valleytronic physics with bosonic condensates. We further believe that our work paves the way towards highly efficient, ultra-compact polariton-based light sources and valleytronic devices based on bosonic quantum fluids hosted in atomically thin materials, which ultimately can be operated at room temperature.

Correspondence and requests for materials should be addressed to Christian Schneider (christian.schneider@physik.uni-wuerzburg.de).

**Acknowledgements**

We acknowledge fruitful discussions and support on the experiment by Mike Fraser. C.S. acknowledges support by the ERC (Project unLiMIt-2D), and the DFG within the Project SCHN1376 3-1. The Würzburg group acknowledges support by the State of Bavaria. A.N. and E.C. acknowledge the support from the megagrant 14.Y26.31.0015 and Goszadanie no. 3.2614.2017/4.6 of the Ministry of Education and Science of Russian Federation. A.V.K. acknowledges the support from the St-Petersburg State University in framework of the project 11.34.2.2012. S.H. and A.V.K. are grateful for funding received within the EPSRC Hybrid Polaritonics programme grant (EP/M025330/1).

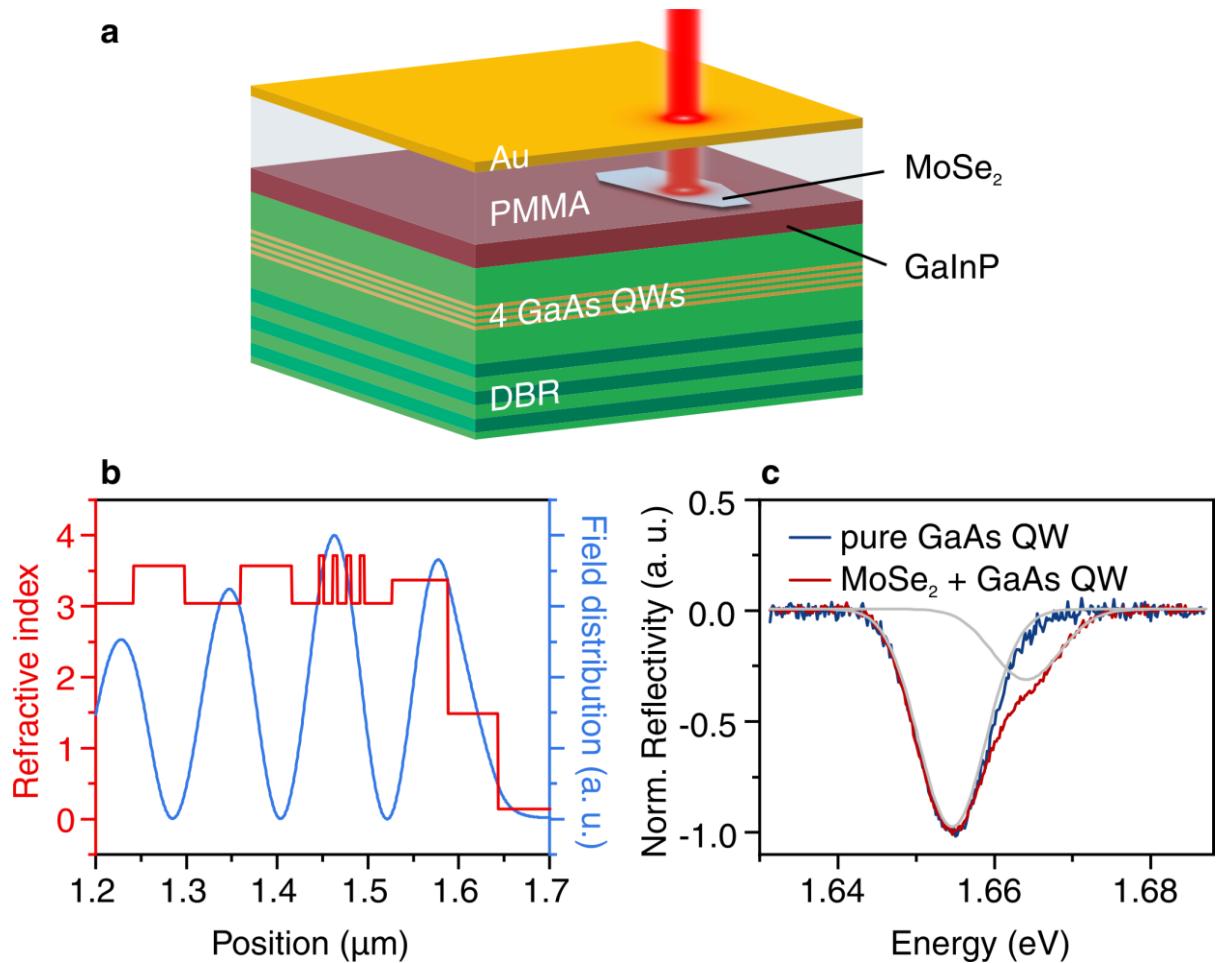

**Fig. 1 | Hybrid Tamm-Monolayer device. a,** Schematic illustration of the Tamm-plasmon device with the embedded MoSe$_2$ monolayer. The monolayer is capped with PMMA, whose thickness primarily determines the frequency of the device's optical resonance. **b,** Calculation of the electromagnetic field intensity in the heterostructure. The field distribution in the device is designed to yield optimal overlap with the position of the quantum wells as well as the atomic monolayer. **c,** Reflectivity spectrum of the device prior to capping the structure with Au. The two absorption dips are correlated with the GaAs-exciton and the MoSe$_2$ neutral excitonic transition.

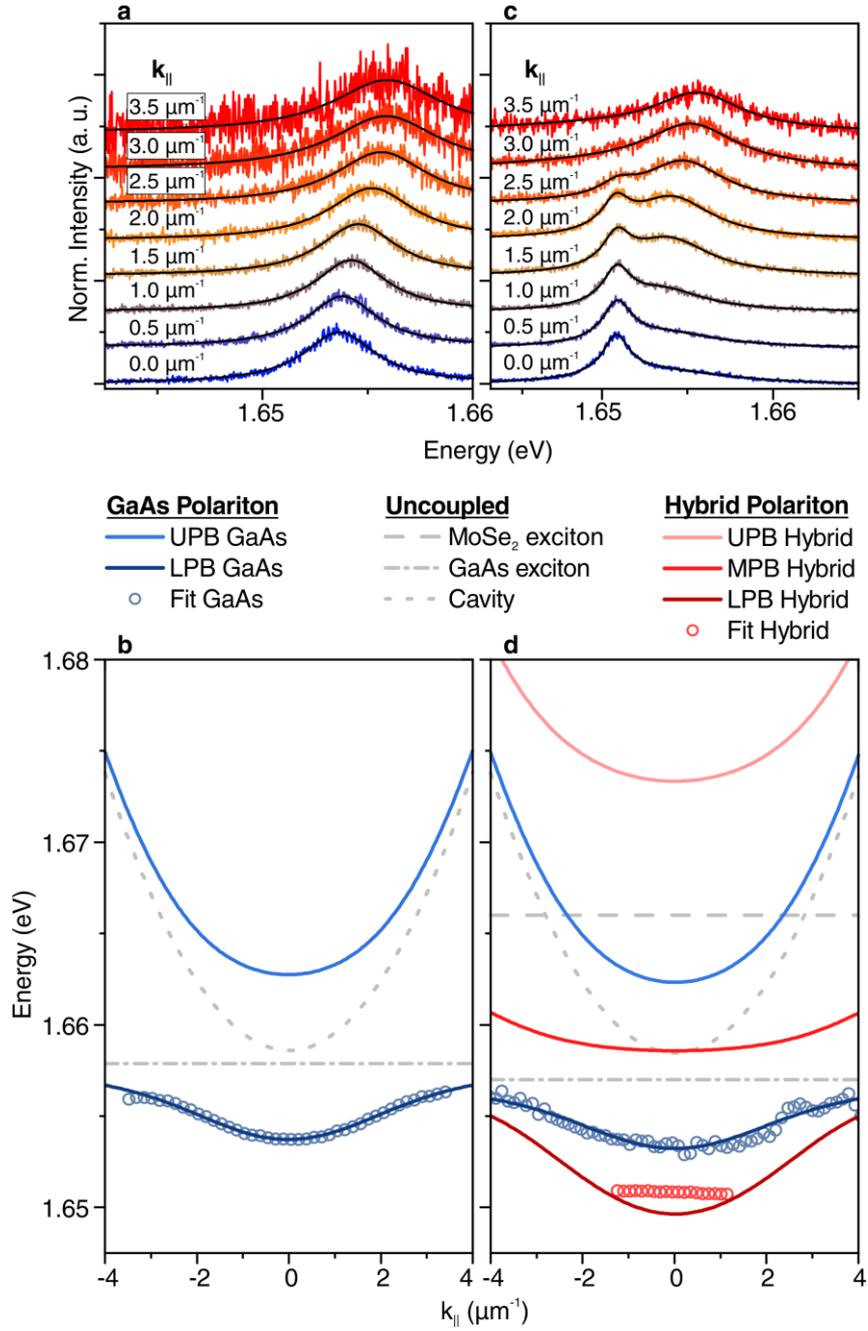

**Fig. 2 | Luminescence of GaAs-exciton-polaritons and hybrid polaritons. a,** Momentum resolved photoluminescence spectra recorded from the device at 4.2 K at the periphery of the monolayer, depicted in a waterfall representation. **b,** Energy-momentum dispersion relation of the signal, following the model of the lower polariton branch in a coupled oscillator system. **c,** Plot of the photoluminescence from the device at the $MoSe_2$ monolayer position. Two peaks evolve, which are attributed to the hybrid polariton mode and the GaAs-polariton resonance from the surroundings of the monolayer. **d,** Energy-momentum dispersion relation of the two modes corresponding with the signals shown in panel **c**. The discrete, hybrid polariton mode is a result of an admixture of 21.0 % GaAs, 21.5 % $MoSe_2$ and 57.6 % photon (appendix).

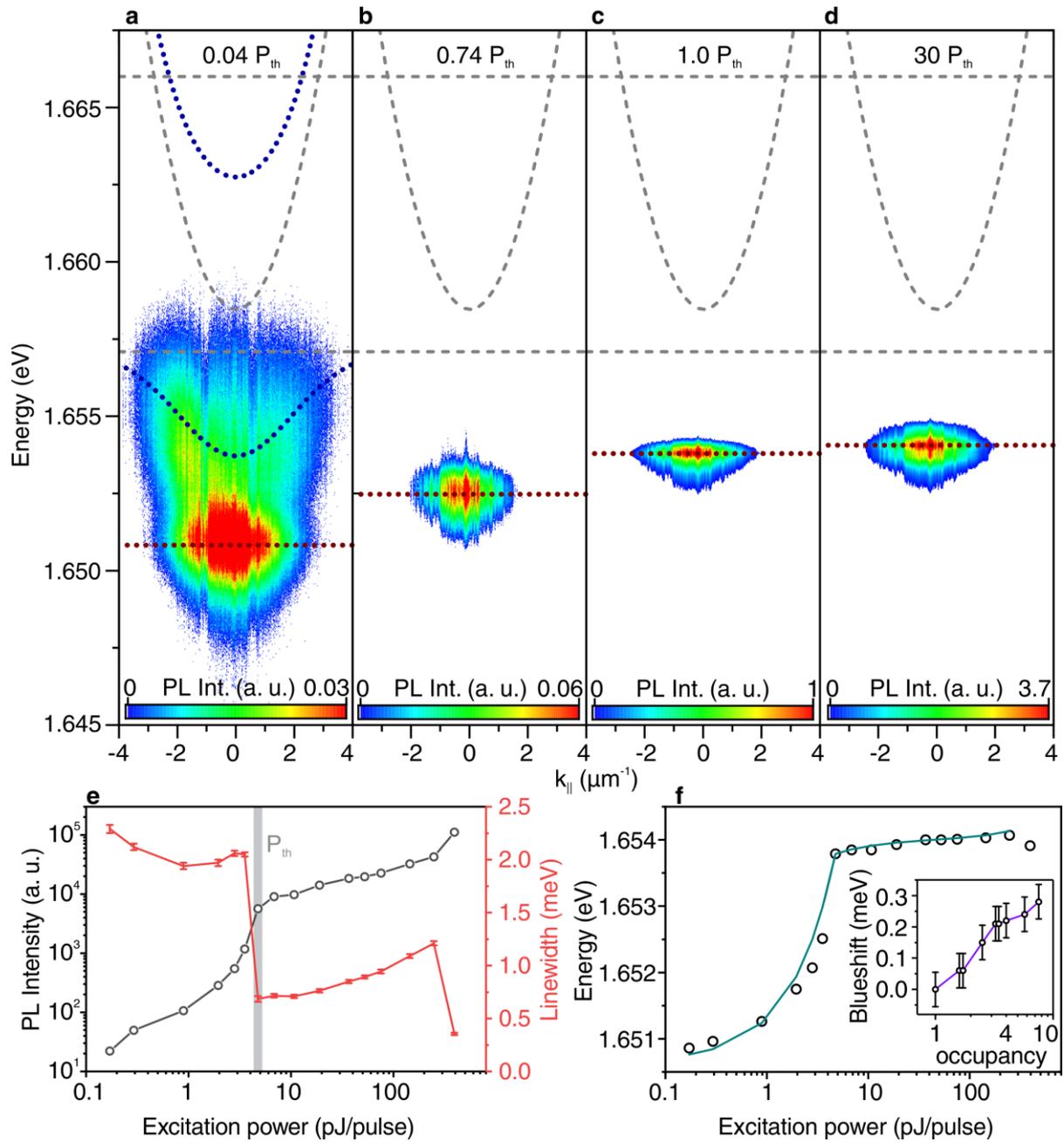

**Fig. 3 | Density-dependent characterization of the hybrid exciton-polaritons. a-d,** False color intensity profile of the hybrid polariton device at different excitation powers, below (**a**), close to (**b**), at (**c**), and (**d**) above the threshold. The red dotted lines resemble the energy of the mode and serve to illustrate the blueshift at increasing excitation powers. **e,** PL emission intensity (black) and linewidth (red) as a function of the excitation power. **f,** Blueshift of the hybrid polariton mode across the condensation threshold. Inset: Blueshift above the threshold as a function of the polariton occupancy, which was normalized to unity at the threshold.

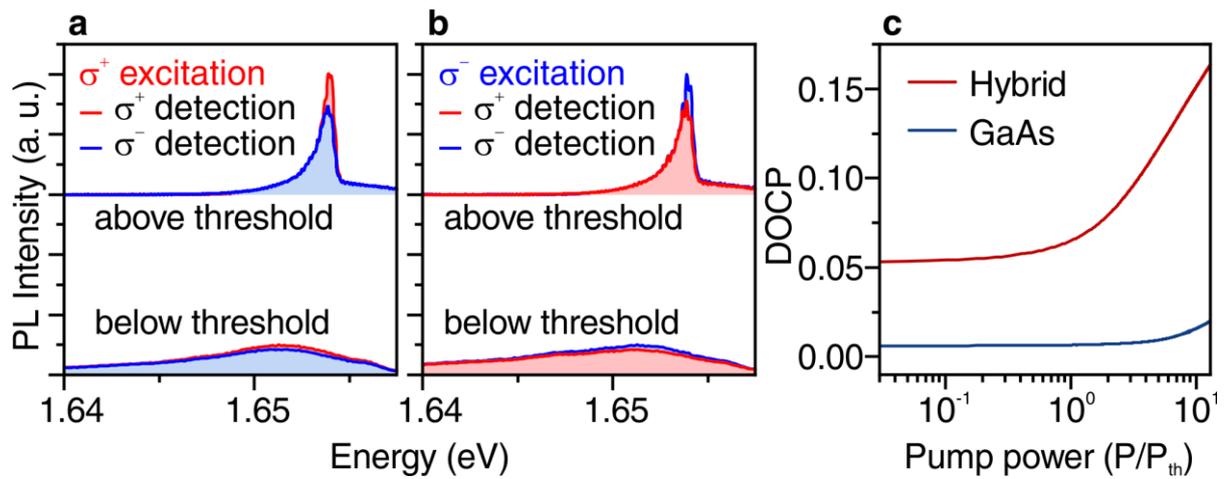

**Fig. 4 | Polarized emission from the hybrid condensate. a,b,** Polarization resolved spectra at k=0 at a pump power of 10*$P_{th}$ above and below the laser threshold for $\sigma^+$ (**a**) and $\sigma^-$ (**b**) excitation. The retained degree of circular polarization is 17.9 % and 16.4 %, respectively. In contrast, below the threshold only 9.8 % ($\sigma^+$) and 7.2 % ($\sigma^-$) are conserved. **c,** Calculation of the pump power evolution of the degree of circular polarization of hybrid polaritons (red) and GaAs-exciton (blue) polaritons.

# Appendix

## section 1

In order to describe the eigenenergies of the hybrid polariton resonances, we apply a coupled oscillator model, which reads in the case of three oscillators:

$$\begin{bmatrix} E_{ex_1} & 0 & V_1/2 \\ 0 & E_{ex_2} & V_2/2 \\ V_1/2 & V_2/2 & E_{cav} \end{bmatrix} \begin{bmatrix} X_1 \\ X_2 \\ C \end{bmatrix} = E \begin{bmatrix} X_1 \\ X_2 \\ C \end{bmatrix}$$

the three Hopfield coefficients quantify the admixture of QW- and monolayer-exciton ($|X_1|^2$; $|X_2|^2$) and cavity photon $|C|^2$. Solving the eigenvalue problem yields the characteristic dispersion relation of hybrid polaritons, featuring three polariton branches. $E_{cav}$ and $E_{ex}$ are photon and exciton energies, respectively, and $V_i$ the exciton-photon coupling strength for the respective oscillator.

## section 2

Here, we provide details on a reference study of the power dependent behavior of the photoluminescence from pure GaAs-exciton polaritons emerging in the periphery of the monolayer device with a detuning of -5.9 meV. Sample conditions are comparable to the main text with a temperature of 5 K and optical excitation with an 82 MHz repetition rate, 2 ps pulsed Ti:Sa laser, tuned to an energy of 1.6732 eV. In Fig. S1a-c we plot the characteristic far field spectra at injection energies of 2.4, 12 and 341 pJ/pulse. From the energy momentum dispersion relations, we can extract the energy, linewidth and the intensity input-output characteristics (see Fig. S1), by integrating the ground state from k = -0.1 µm⁻¹ to 0.1 µm⁻¹ and fitting it with a Lorentz function. At an approximate average pump power of 182 pJ/pulse, we observe a distinct super linearity in the emission intensity in panel d, with a threshold-like s-shape indicative for a laser-like transition. The emission energy of the studied mode after threshold is approximately pinned to the empty cavity resonance [see panels c and f]. This feature is typically attributed to a transition from the strong- to the weak coupling regime, which typically occurs at or slightly above the Mott density in high-quality

QW-microcavities. The input power approximately corresponds to the second threshold observed in the main text, attributed to the Mott-transition of the bare GaAs QW directly in the vicinity to the monolayer.

**section 3**

This section compares the polarization of the emission from our hybrid monolayer-GaAs device with the emission from pure GaAs polaritons, recorded at a comparable laser-lower polariton detuning in relation to the hybrid polaritons of Fig. 4 (main text). The experiment was carried out under the same experimental conditions as described in the main text for Fig 4.

In Fig. S2 we plot the resulting degree of circular polarization (DOCP) from the emission of the cavity, both subject to $\sigma^+$ and $\sigma^-$ pumping. We note, that the DOCP does not show any distinct power dependency within the error margin, as expected in the linear regime by our Boltzmann model, and is significantly lower than the DOCP which we have record from our hybrid polariton condensate under significantly lower pump powers.

**section 4**

We provide additional details on the blueshift fit in Fig. 3f of the main text. The emission from the polariton condensate is subject to an energy shift, depending on the pumping power, both below and above the condensation threshold. At low pumping powers, the occupation of the ground state is negligible and the energy shift arises from the polariton interaction with the excitonic reservoir which builds up with increased excitation power. In the conventional single reservoir model[1] the homogeneous reservoir density is fixed above the threshold due to the stimulated scattering into the ground state. The repulsive polariton-polariton interactions thus govern the further increase of the emission energy above the threshold, which occurs at a reduced rate compared to the large below-threshold blueshift due to polariton-exciton interactions. This model, despite having the advantage of simplicity, is not sufficient for the description of the hybrid polariton devices, where the emission blueshift depends nonlinearly on the pumping power above the lasing threshold.

To account for the hybrid nature of the microcavity we solve the coupled rate equations for the occupations of the polariton condensate mode and the two exciton reservoirs, neglecting the spin polarization and spontaneous scattering probability:supp

$$\frac{dN}{dt} = \left(W_1 n_1 + W_2 n_2 - \frac{1}{\tau}\right) N, \tag{S1}$$

$$\frac{dn_i}{dt} = P_i - \left(W_i N + \frac{1}{\tau_i}\right) n_i. \tag{S2}$$

Here, $N$ is the condensate occupation, $n_1$ and $n_2$ are the densities of exciton reservoirs in the QW and the TMDC monolayer, respectively, $W_1$ and $W_2$ are the corresponding scattering rates, $\tau$ and $\tau_i$ are the condensate and reservoir lifetimes, and $P_i = P g_i$ are the reservoir pumping rates. Equating the time derivatives to zero we find the stationary values of $N$ and $n_i$.

Below the condensation threshold threshold ($N = 0$) we express the reservoir densities $n_i = P g_i \tau_i$ from Eq. (S2). Above the threshold ($N > 0$) both reservoir densities $n_i = P g_i / (W_i N + \tau_i^{-1})$ are depleted by stimulated scattering into the condensate. The occupation of the latter reads:

$$N = \frac{N_1 + N_2}{2} + \sqrt{\left(\frac{N_1 - N_2}{2}\right)^2 + P^2 \tau^2 g_1 g_2}, \tag{S3}$$

where $N_i = P g_i \tau - W_i^{-1} \tau_i^{-1}$ has the physical meaning of the occupation of a condensate, which is only coupled to the $i$-th reservoir. The total blueshift is governed by the polariton condensate interaction with itself and with the two reservoirs as follows:

$$\Delta E = (\alpha_1 |X_1|^4 + \alpha_2 |X_2|^4) N + \alpha_1 |X_1|^2 n_1 + \alpha_2 |X_2|^2 n_2, \tag{S4}$$

where $X_1$ and $X_2$ are the exciton Hopfield coefficients of the polariton mode, corresponding to the QW and TMDC exciton, respectively.

The fits of the experimental dependence of blueshift on the pumping power, obtained with Eqs. (S1, S2) and Eq. (S4) are shown in Fig. 3f. The initial nonlinear increase in the blueshift above the threshold corresponds to a redistribution of the excitonic densities between the two reservoirs, so that the total condensate gain provided by stimulated scattering remains equal to the radiative losses. Note that the two reservoirs model captures the nonlinear behavior of blueshift more accurately than the conventional single reservoir model[1], demonstrating the important role of the interplay of the two reservoirs in the behavior of the hybrid polariton laser.

The following realistic parameters were used in the fit: $\tau = 0.5$ ps, $\tau_1 = 500$ ps, $\tau_2 = 10$ ps. We also assume equal excitonic interaction constants[2] $\alpha_1 = \alpha_2$ and take equal Hopfield coefficients $X_1 = X_2$ from the polariton branch fit in Fig. 2 of the main text. Note that, under this assumptions, the reservoir contribution to the condensate blueshift is proportional to the total reservoir density $n = n_1 + n_2$. The free parameters used for the fit are the two relations between the pumping efficiencies $g_1/g_2 = 1$ and the scattering rates $W_1/W_2 = 0.02$.

The details of the fit presented in Fig. 3f are shown in Fig. S3. Below the threshold, the blueshift is linear in the pumping power. The population of QW excitons builds faster than in the TMDC due to longer exciton lifetime in the QW. Above the threshold, both reservoirs are depleted by stimulated scattering into the macroscopically populated polariton mode. Depletion efficiency is governed by the spontaneous scattering rate. Stronger coupling of the TMDC excitonic mode to the polaritonic mode favors stronger TMDC reservoir depletion and results in a slight decrease of its occupation above the threshold. In contrast, the QW reservoir grows so that the total gain provided by the two reservoirs compensates the radiative losses of the condensate. This results in a nonlinear dependence of the blueshift on the pumping power above the threshold. At higher powers, however, the combined reservoir density reaches a steady state, and the blueshift becomes linear because it is governed by polariton-polariton interactions as in the single-reservoir case[1].

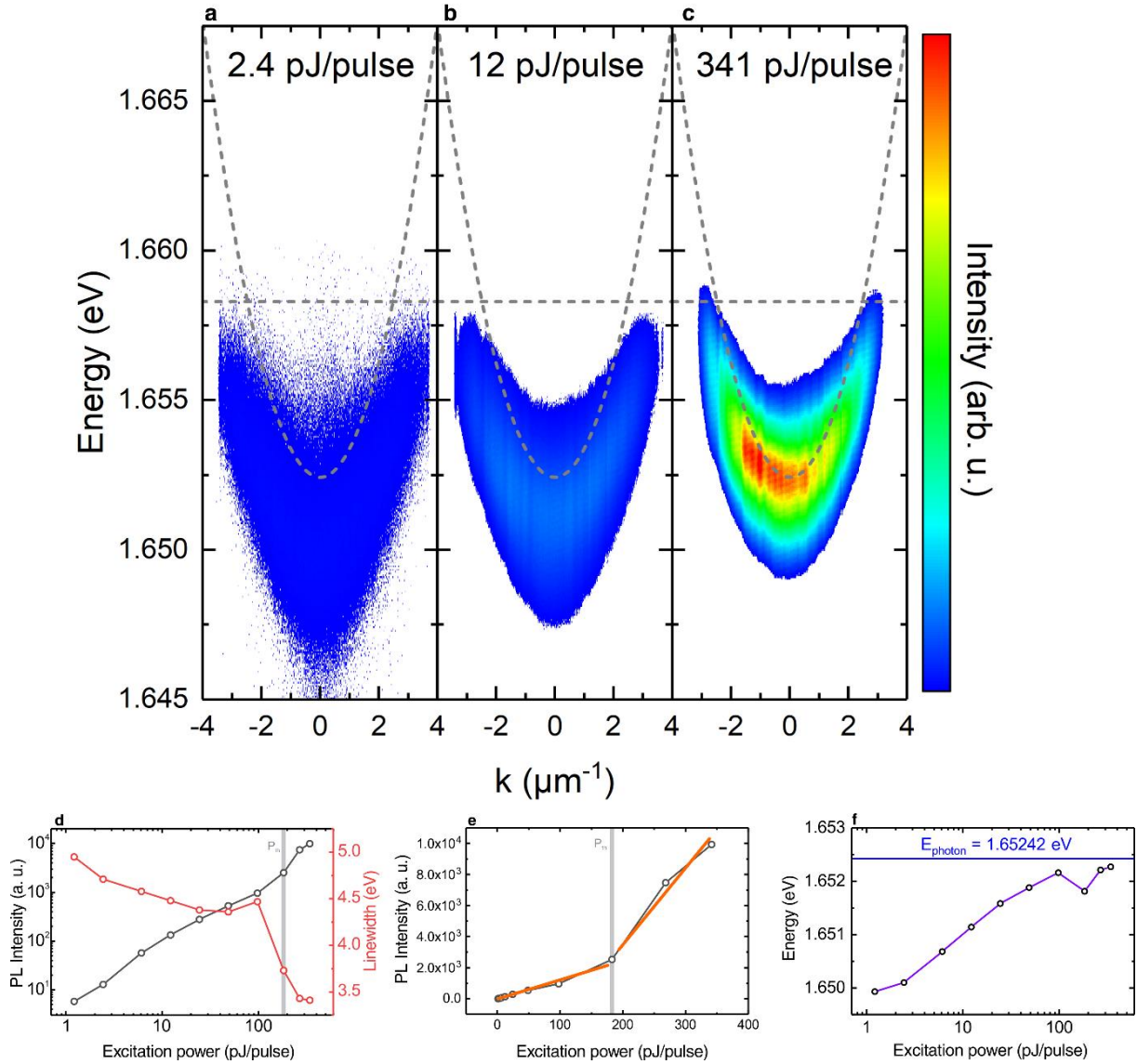

**Fig. S1 | False color intensity profile of the pure GaAs device at a detuning of −5.9 meV at different excitation powers. a,b,** Lower polariton emission under the threshold (2.4 and 12 pJ/pulse, respectively) and **c,** photon lasing above the threshold (341 pJ/pulse). The flat dashed line (parabolic dashed line) corresponds to the exciton mode (photon mode). **d,** PL intensity (black) with the distinct s-shape and linewidth (red) as a function of excitation power. In contrast to the hybrid device, we observe only one threshold at much higher pump powers, corresponding to the conventional inversion based lasing transition. **e,** PL intensity as a function of excitation power on a linear scale. **f,** PL emission energy (blue) as a function of the excitation power.

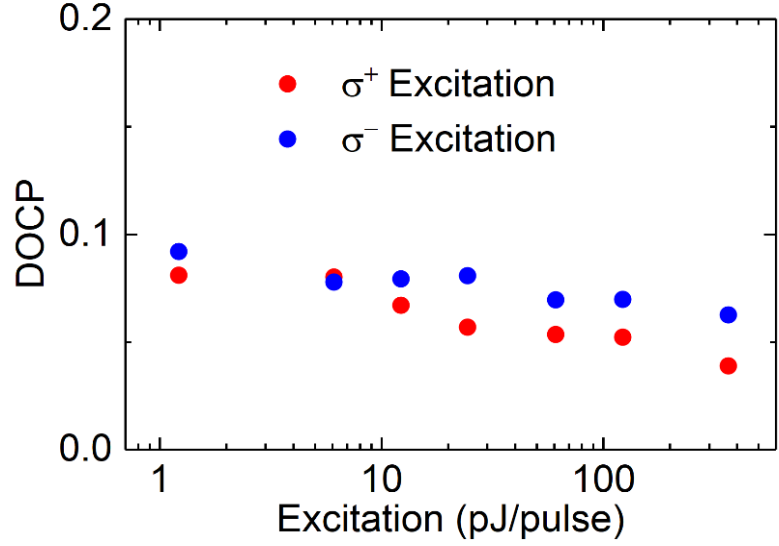

**Fig. S2 |** Degree of circular polarization from the pure GaAs lower polariton branch for both $\sigma^+$ (red) and $\sigma^-$ (black) polarized excitation for comparable laser-LP detuning (in relation to the hybrid GaAs-MoSe$_2$ LP) depending on excitation power. We observe no increase even for strong laser pump and the DOCP is significantly lower than in the condensate phase of the hybrid GaAs-MoSe$_2$ device displayed in the main text, which substantiates the claim regarding polarization conservation due to bosonic condensation in a hybrid MoSe$_2$-GaAs microcavity.

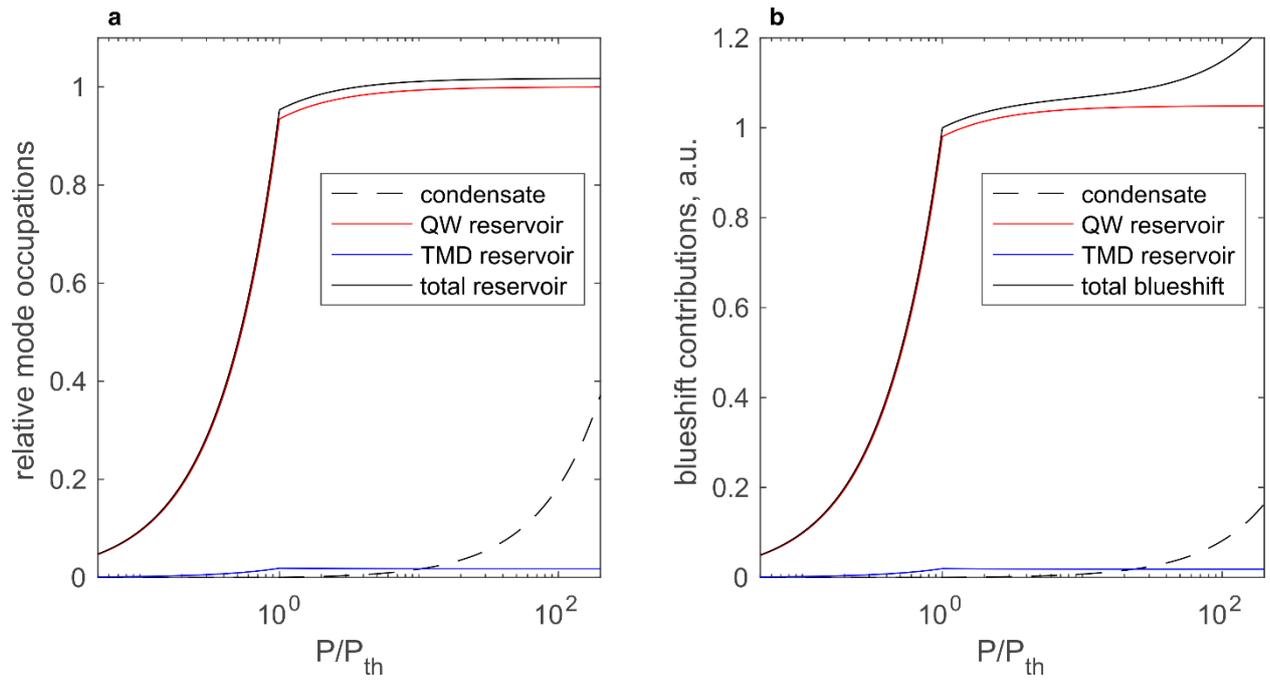

**Fig. S3 | a,** Relative polariton condensate and reservoir occupations. **b,** Relative contributions to the total emission blueshift from the condensate self-interaction and the two reservoirs.